\documentclass{article}

\usepackage{color}
\usepackage{latexsym}
\usepackage{amssymb}
\usepackage{gensymb}
\usepackage{subfigure} 
\usepackage{booktabs}
\usepackage{diagbox}
\usepackage{multirow}
\usepackage{array}
\usepackage{threeparttable}
\usepackage{hyperref}
\hypersetup{hidelinks}
\usepackage[utf8]{inputenc}
\usepackage{cite}
\usepackage{graphicx}

\usepackage{geometry}
\usepackage{amsmath}

\usepackage{algorithmic}

\usepackage{array}

\usepackage{authblk}
\usepackage{textcomp}
\usepackage{color}
\usepackage[misc]{ifsym} 

\begin{document}

\title{Theoretical Conditions for Field-free Magnetization Switching Induced by Spin-orbit Torque and Dzyaloshinskii-Moriya Interaction}
\author[1,2]{Min Wang}
\author[1,2,3,*]{Zhaohao Wang }
\author[1,2,3]{Xueying Zhang}
\author[1,2,3,*]{Weisheng Zhao }
\affil[1]{Fert Beijing Research Institute, Beihang University, Beijing, China}
\affil[2]{School of Microelectronics, Beihang University, Beijing, China}
\affil[3]{Beijing Advanced Innovation Center for Big Data and Brain Computing, Beihang University, Beijing, China}
\affil[*]{zhaohao.wang@buaa.edu.cn and weisheng.zhao@buaa.edu.cn}
\date{}
\maketitle

\begin{abstract}
Recently, it was demonstrated that field-free switching could be achieved by combining spin-orbit torque (SOT) and Dzyaloshinskii-Moriya interaction (DMI). However, this mechanism only occurs under certain conditions which have not been well discussed. In this letter, it is found that the ratio of domain wall width to diameter of nanodots could be utilized as a criteria for judging this mechanism. Influences of different magnetic parameters are studied, including exchange constant, DMI magnitude and field-like toque, etc. Besides, we reveal the importance of the shrinkage of magnetic domain wall surface energy for metastable states. Our work provides guidelines to the experiments on the DMI-induced field-free magnetization switching, and also offers a new way for the design of SOT-based memory or logic circuits.


\end{abstract}

\section{Introduction}

Spin-orbit torque (SOT)-induced switching is considered the next-generation write technology for spintronic memory, owing to its ultrafast speed and low power consumption\cite{ramaswamy2018recent,liu2012spin,garello2014ultrafast,wang2018high}. Concerning on the drawback of SOT, i.e., the need of external magnetic field, numerous efforts have been devoted to searching for field-free SOT schema \cite{yu2014switching,wang2018field,fukami2016magnetization,yang2019spin}. Recently, it was demonstrated that, in the case of non-zero Dzyaloshinskii-Moriya interaction (DMI), the perpendicular magnetization can be efficiently switched through SOT current-induced domain nucleation and domain wall (DW) motion without the need of magnetic field\cite{chen2019field,wu2020deterministic,liu2020anomalous}. However, the origin of this mechanism and the theoretical conditions have been less discussed. 

Figure 1(a) presents the studied device  which uses a common SOT multilayer structure without additional complexity. Figure 1(b)-(d) show typical magnetization dynamics in free layer via SOT and DMI in the absence of external magnetic field. All these results are obtained by performing micromagnetic simulation with OOMMF package\cite{donahue1999oommf}. As shown in Fig. 1(b)-(c), after the nucleation at the edge, DW moves across the midline and causes non-zero z-component magnetization, which means deterministic switching and following is defined as $bias$-$state$. Nevertheless, the occurrence of $bias$-$state$ is conditional. By just changing the exchange constant in Fig. 1(d), the DW moves and stays at the midline, leading to non-deterministic switching. The reasons behind it need to be discussed, which is the main focus of this letter.



\begin{table}[h]
	\renewcommand{\arraystretch}{1.3}
	\caption{Parameters of simulation }
	\label{table1}
	\centering
	\begin{tabular*}{0.8\textwidth}{@{\extracolsep{\fill}}lll}
		\hline
		Symbol & Parameter & Default value \\
		\hline
		-       & MTJ area                 &$0.25\times\pi\times40~{\rm nm}\times40~{\rm nm}$\\
		$t_F$   &Thickness of FL           &$1.0~{\rm nm}$\\
		$\alpha$&Damping constant          &$0.1$ \\
		$A_{ex}$     & Exchange constant        & $1\times10^{-11}~{\rm J/m}$ \\
		$K_u$   & Magnetic anisotropy      & $8\times10^5~{\rm J/m^3}$\\
		$M_s$   & Saturation magnetization &$1\times10^6~{\rm A/m}$\\
		$D$     & DMI magnitude            &$0.5~{\rm mJ/m^2}$\\
		$\theta_{SOT}$&Spin Hall angle     &$0.3$ \\	
		\hline
	\end{tabular*}
\end{table}


Following we define $m_{z,f}$ as final-state z-component magnetization $<$$m_z$$>$ which is formed at the end of current pulse. The extreme value amongst a group of $m_{z,f}$ caused by the various current densities is defined $m_{z,ex}$, which judges whether or not $bias$-$state$ occurs. The parameters and their default values are listed in Table I, unless otherwise stated. In the rest of this letter, we will discuss the theoretical conditions for the deterministic switching induced by SOT and DMI, through analysing the DW width, DW surface energy, and shape effect.

\section{Theoretical conditions}
\subsection{$K_u$, $A_{ex}$, $M_s$}

Figure 2(a) shows the phase diagram of $m_{z,ex}$ under various $A_{ex}$ and $K_u$, where the area outside the dark red indicates the existence of $bias$-$state$. The dotted line in Fig. 2(a) represents the borderline between deterministic and non-deterministic switching, and the fitting curve of the borderline is expressed as:
\begin{equation}
K_{eff}=K_u-0.5\mu_0(N_z-N_x)M_s^2=\frac{A_{ex}-d_1}{(diameter.d_2)^2}
\label{fit}
\end{equation}
where $N_i$ is the demagnetization factor of the film in the $i$ direction. $d_1$ and $d_2$ are fitting values, $d_1$=$0.239\times10^{-11}~{\rm J/m}$, $d_2$=0.169.

Equation \eqref{fit} indicates that the smaller $A_{ex}$ requires a lower $K_{eff}$ to induce the deterministic switching, which helps in decreasing the critical switching current. However, excessively small $A_{ex}$ could reduce the DW energy so that the magnetic domains are chaotic. In addition, with the increasing $K_u$ or decreasing $A_{ex}$, $|m_{z,ex}|$ monotonically increases and thus the reliability of magnetization switching is improved.

The borderline concerning $K_u$ and $M_s$ is presented in Fig. 2(b). Below the green line in Fig. 2(b), $m_{z,ex}<-0.01$, indicating the occurrence of $bias$-$state$. Conversely, DW is driven and finally stays at the midline so that the switching becomes non-deterministic. A good agreement between simulation results and \eqref{fit} is achieved. The above results show that the switching behaviours of the device could be adjusted by changing $K_u$, which could be implemented with voltage-controlled magnetic anisotropy (VCMA) effect.

From the above analysis, the switching process depends on the  $A_{ex}$, $K_u$ and $M_s$, which could be together taken into account by the expression of DW width, as follows\cite{thiaville2012dynamics,nasseri2018collective}:

\begin{equation}
\label{eqDW}
\Delta=\sqrt{\frac{A_{ex}}{K_{eff}}}=\sqrt{\frac{A_{ex}}{K_u-0.5\mu_0(N_z-N_x)M_s^2}}
\end{equation}

Therefore, $\Delta$ could be considered a characteristic parameter reflecting the behaviors of magnetization switching. In order to obtain the $bias$-$state$, $\Delta$ should be lower than a critical value. 

\subsection{D and field-like torque}
As shown in Fig. 3(a), the DMI magnitude has a dramatic influence on the magnetization dynamics. Larger $K_u$ is required for generating the $bias$-$state$, which is consistent with \eqref{eqDW}. It is seen that the critical $K_u$ for the deterministic switching increases with the enhanced DMI when $D$ is less than 0.5$~{\rm mJ/m^2}$, where the texture of DW  evolves from  Bloch-like type to Néel-like type (see Fig. 3(c)). However, the critical $K_u$ becomes stable when $D$ is in the range of  0.5$\sim$1.0$~{\rm mJ/m^2}$. When $D$ is smaller, the $bias$-$state$ is prone to occur, wheareas $|m_{z,ex}|$ is smaller. This means a trade-off between the DMI effect and reliability of magnetization switching.

Here we define damping-like torque and field-like torque as
\begin{equation}
\boldsymbol{\tau_{DL}}=J_{SOT}\xi\lambda_{DL}\mathbf{m}\times({\boldsymbol{\sigma}}\times\mathbf{m})
\label{DL}
\end{equation} 
\begin{equation}
\boldsymbol{\tau_{FL}}=J_{SOT}\xi\lambda_{FL}({\boldsymbol{\sigma}}\times \mathbf{m})
\label{FL}
\end{equation}
where $\xi=\gamma\hbar/(2et_FM_s)$. $\lambda_{DL}$ and $\lambda_{FL}$ represent the magnitudes of damping-like torque and field-like torque, respectively. 

Recently, experiments have demonstrated that the ratio of field-like torque to damping-like torque can vary at a wide range\cite{kim2013layer,avci2014fieldlike,qiu2014angular}. The switching behaviours are analysed with various $\lambda_{FL}/\lambda_{DL}$ in the range of -0.5 to 0.5 (in our convention as \eqref{DL} and \eqref{FL}). Firstly, it is seen from Fig. 4(a)-(c) that the positive and negative $\lambda_{FL}/\lambda_{DL}$ could, respectively, increase and decrease the critical nucleation current density. Second, positive $\lambda_{FL}/\lambda_{DL}$ prohibits the generation of $bias$-$state$ and requires lower critical $\Delta$ (see Fig. 4(b) and (d)), whereas the existence of negative $\lambda_{FL}/\lambda_{DL}$ increases the critical $\Delta$ (see Fig. 4(d)). Note that the abnormal region in the bottom right corner of Fig. 4(a) could be attributed to the improper nucleation position.

To induce deterministic field-free switching in a 40 nm circle film, recommended guideline for different parameters is summarized in Table \ref{table2}. Higher $K_u$ is preferable, whereas $A_{ex}$, $M_s$ and $\lambda_{FL}/\lambda_{DL}$ need to be lower, to obtain a $\Delta$ lower than critical value. Smaller $D$ is beneficial to the occurrence of $bias$-$state$ whereas larger $D$ leads to larger $|m_{z,ex}|$. 

\begin{table}[h]
	\renewcommand{\arraystretch}{1.3}
	\caption{Summary in nanodots with diameter of 40nm}
	\label{table2}
	\centering
	\begin{tabular*}{0.8\textwidth}{@{\extracolsep{\fill}}ccccccc}
		\hline
		$\Delta$ &$A_{ex}$ &$K_u$&$M_s$  &$D$   &$\lambda_{FL}/\lambda_{DL}$\\
		\hline
		$-$      &$-$ &$+$  &$-$   &$\star$&$-$ \\	
		\hline
	\end{tabular*}
	\begin{tablenotes}
		\footnotesize
		\item[1] $+$ higher; $-$ lower; $\star$ choose the appropriate value.
	\end{tablenotes}
\end{table}

\subsection{Scaling study}
Now diameter of nanodots is taken into account. We define $p$ as the ratio of $\Delta$ to diameter. As shown in Fig. 5(a), the critical $p$ for generating the $bias$-$state$ fluctuates in a small range for various diameters. When $p$ is lower than critical value, $bias$-$state$ occurs under the action of SOT and DMI. Although those critical $p$ are similar, the dynamics of domains differ from size to size (see Fig. 5(b)). While the diameter is larger, DW prefers to be bent and stabilized away from the midline, leading to higher $|m_{z,ex}|$.

As the diameter increases to 60 nm, shown in Fig. 5(c), the nucleation position is transferred from the edge to the centre, which is attributed to the subvolume domain nucleation arising from the increasing film size \cite{sato2011junction,2017Scaling}. After the complex magnetic domain motion, $bias$-$state$ still occurs under appropriate SOT current. The nucleation position could return to the edge by reducing $K_u$, increasing $A_{ex}$, or increasing $D$ (see Fig. 5(d)). These adjustments suppress the subvolume domain nucleation process.

\section{Shape effect}

The magnetization switching caused by Néel-like DW tends to be restrained in square film, compared with circle film (see Fig. 4(a) and Fig. 6(a)). For a circle film, the DW moves across the midline under the inertial effect in order that the DW length is reduced for minimizing the DW surface energy\cite{kittel1950note,saitoh2004current,brigner2019shape}. As a result, the $bias$-$state$ is formed. As to a square film, the DW stays at the midline since it is not inclined to shrink due to the invariable DW length, especially for Néel-like DW (see Fig. 6(c)). In order to further prove the shape effect, we apply a titled current to the square film, as shown in Fig. 6(b) and (d). Remarkably, the $bias$-$state$ occurs since the DW is able to shrink by moving towards the corner. Therefore, it is confirmed that the shrinkage of the DW surface energy plays an important role in the field-free magnetization switching induced by SOT and DMI.

\section{Conclusion}
We have revealed the theoretical conditions for the field-free deterministic switching induced by SOT and DMI. In a circle film, the deterministic switching occurs when the ratio of DW width to the diameter is smaller than a critical value. The detailed domain dynamics is strongly related to the magnetic parameters, device size and shape. In regard to the shape effect, the shrinkage of DW surface energy favours the occurrence of deterministic switching. Our work highlights the importance of DMI in the domain dynamics and helps in the design of high-speed low-power spintronics devices.

\bibliographystyle{IEEEtran}
\bibliography{reference}
	
\newpage

\begin{figure}[h]     
	\includegraphics[width=12cm]{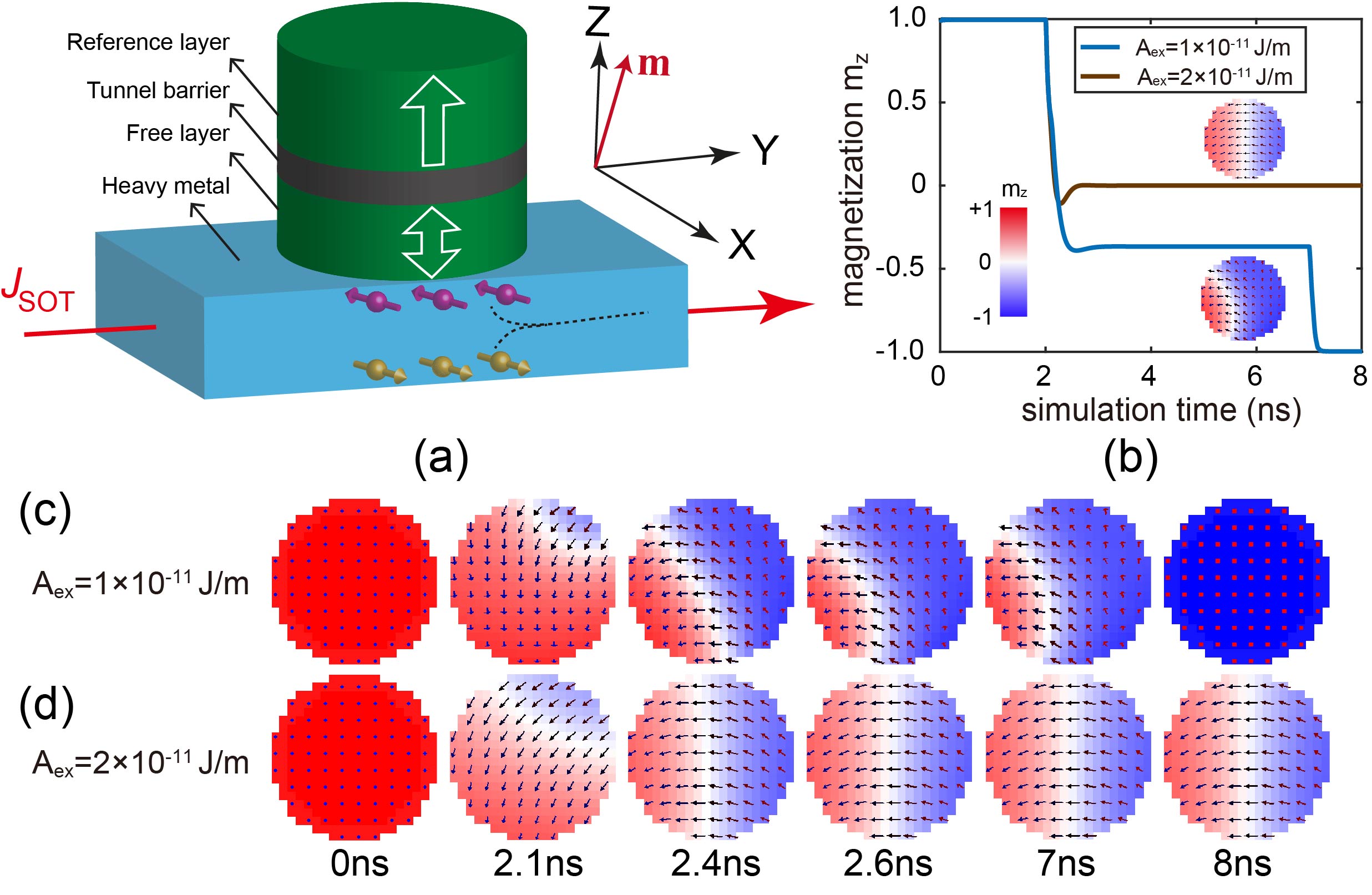}
	\centering
	\label{1}
	\caption{(a) Schematic structure of the studied perpendicular-anisotropy SOT-MTJ. (b) Typical curves of z-component magnetization under the action of SOT and DMI. A 5 ns current pulse is applied after 2 ns relaxation. (c) Snapshots of the magnetization dynamics at various exchange constants. Other parameters are listed in Table I.}	
\end{figure}

\begin{figure}[h]     
	\includegraphics[width=12cm]{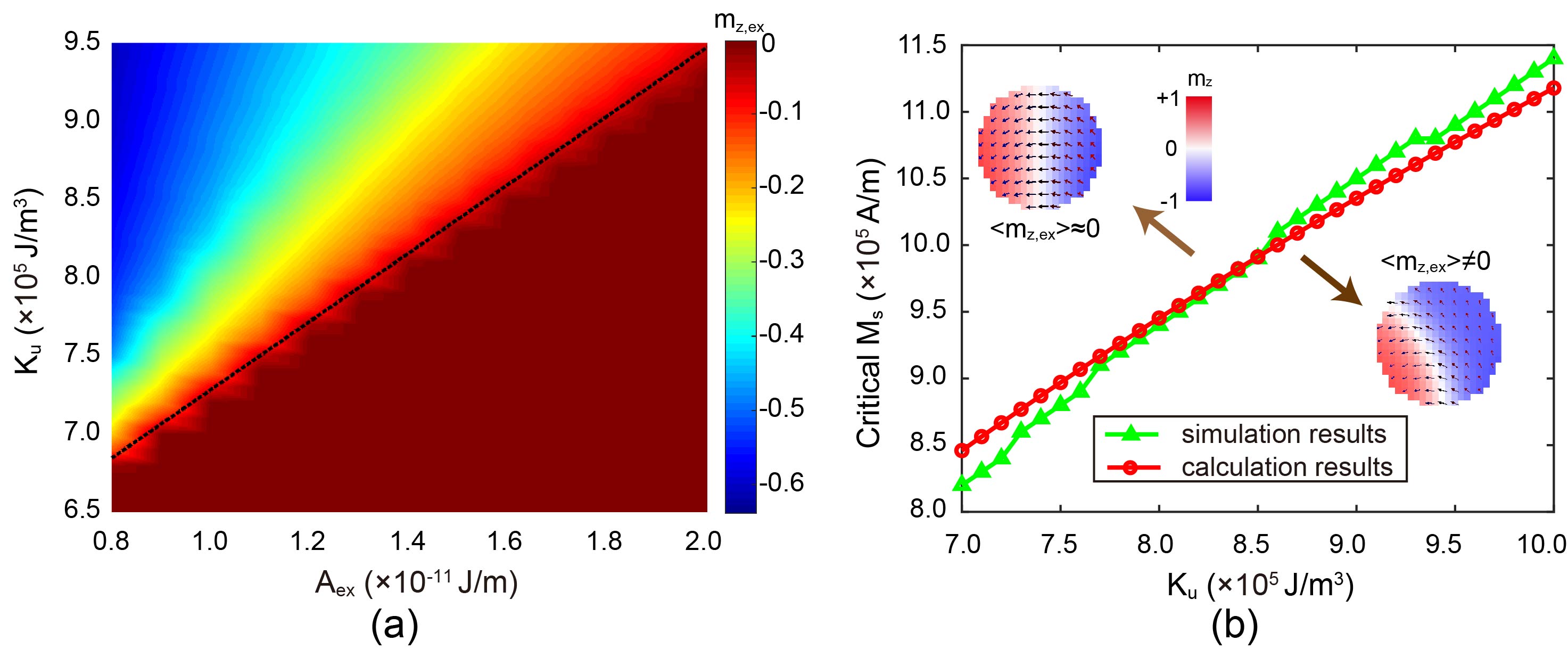}
	\centering
	\label{2}
	\caption{(a) Phase diagram of $m_{z,ex}$ as a function of exchange constant and magnetic anisotropy. The dotted line means the fitting curve of the borderline. (b) Critical $M_s$ for generating the bias-state as a function of $K_u$. $A_{ex}=1.6\times10^{-11}~{\rm J/m}$.}	
\end{figure}

\begin{figure}[h]     
	\includegraphics[width=12cm]{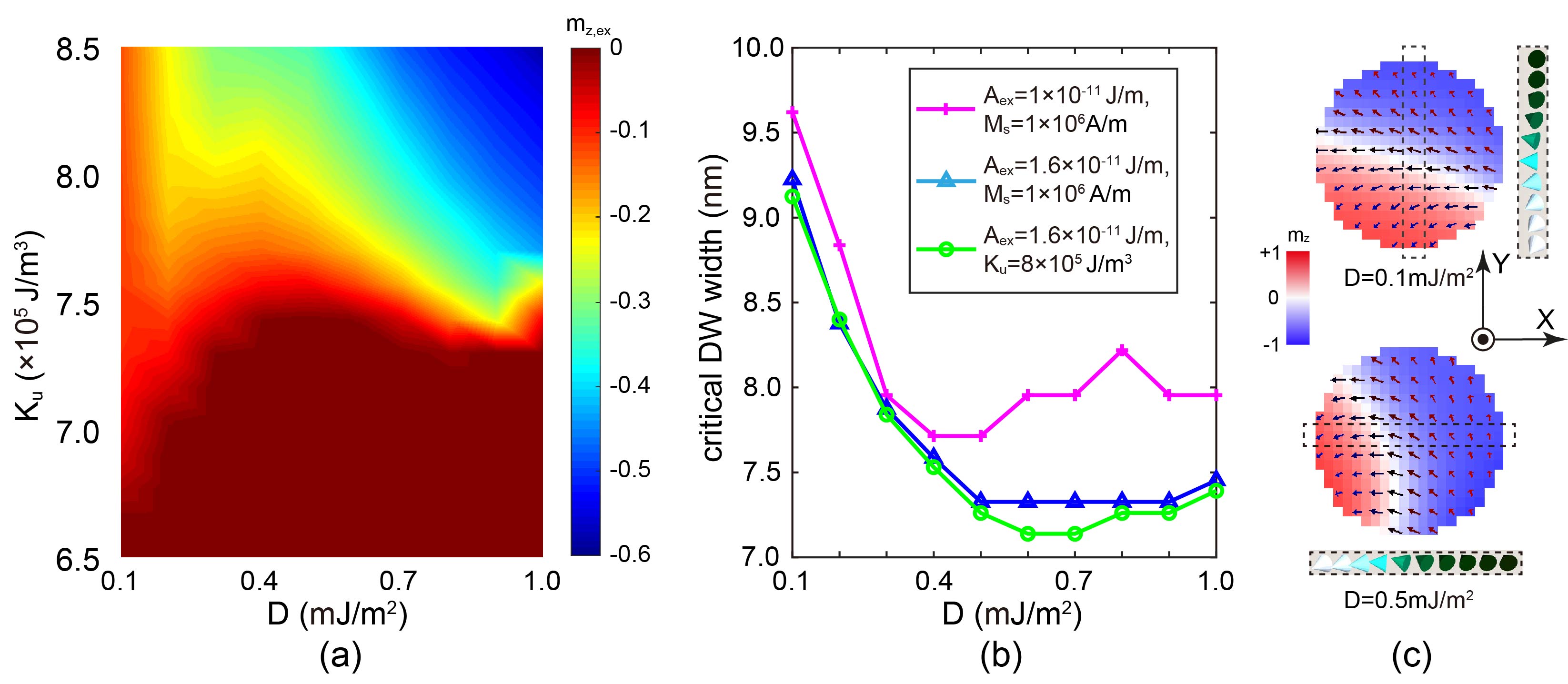}
	\centering
	\label{3}
	\caption{(a) Phase diagram of $m_{z,ex}$ as a function of DMI magnitude and magnetic anisotropy. (b) Critical DW width as a function of DMI magnitude. (c) Snapshots of magnetization of $bias$-$states$ with $D$=$0.1~{\rm mJ/m^2}$ and $D$=$0.5~{\rm mJ/m^2}$.}	
\end{figure}

\begin{figure}[h]     
	\includegraphics[width=12cm]{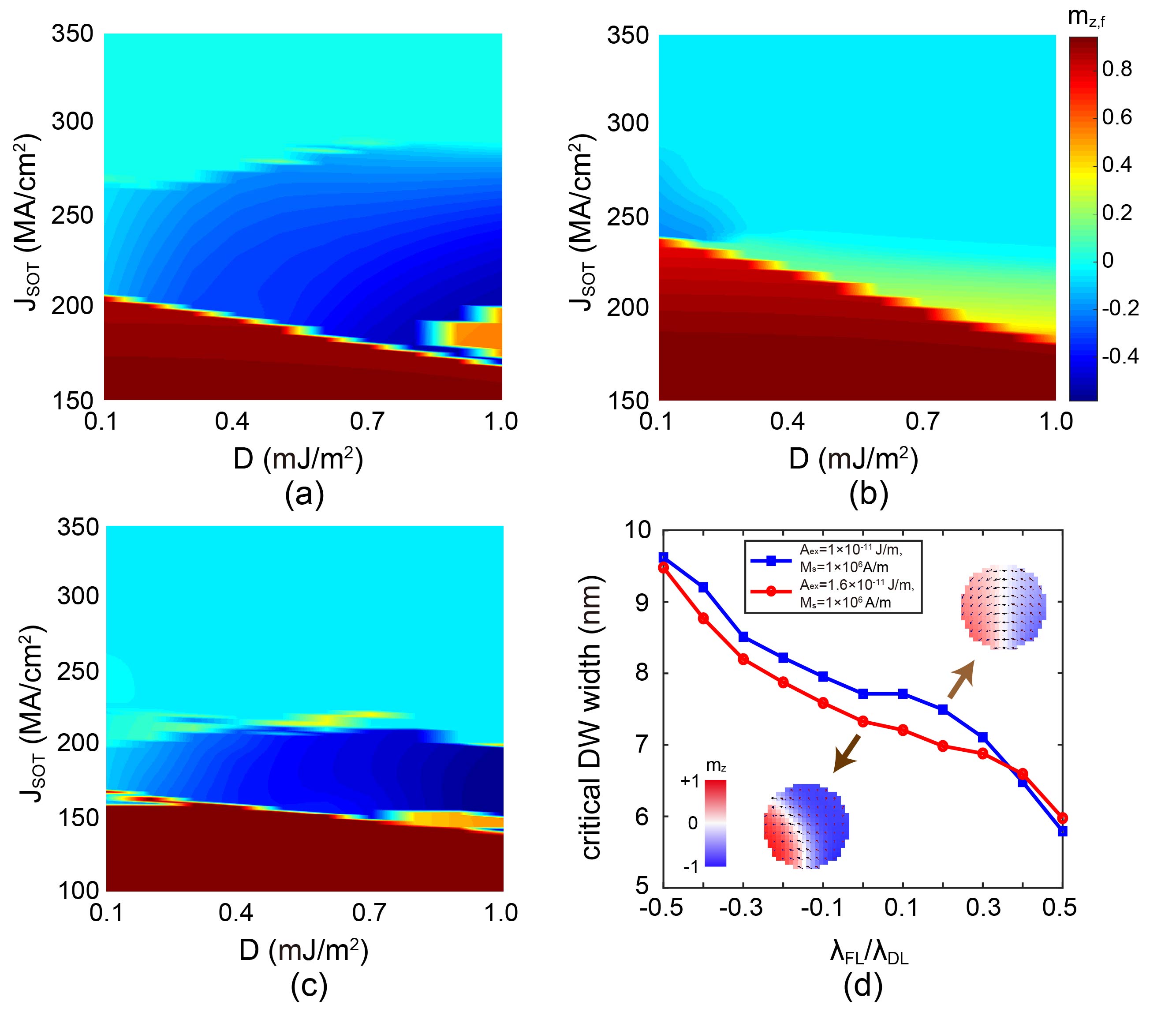}
	\centering
	\label{4}
	\caption{Phase diagrams of $m_{z,f}$ as a function of SOT current density $J_{SOT}$ and DMI magnitude, (a) $\lambda_{FL}/\lambda_{DL}$=$0$, (b) $\lambda_{FL}/\lambda_{DL}$=$0.5$, (c) $\lambda_{FL}/\lambda_{DL}$=$-0.5$. (d) Critical DW width as a function of $\lambda_{FL}/\lambda_{DL}$.}
\end{figure}

\begin{figure}[h]     
	\includegraphics[width=12cm]{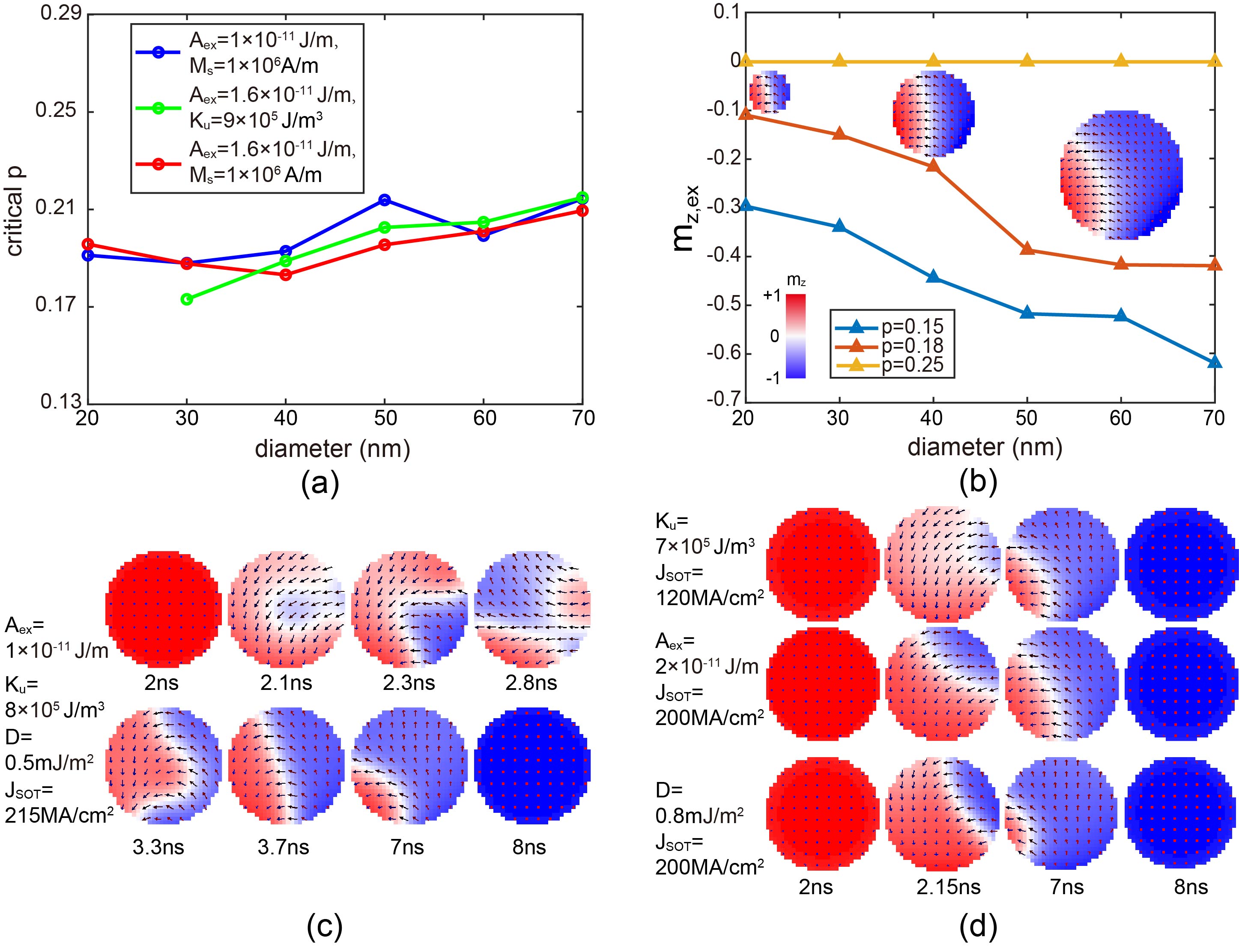}
	\centering
	\label{5}
	\caption{(a) Critical $p$ for generating the $bias$-$state$ as a function of nanodot diameter. (b) The dependence of $m_{z,ex}$ on the nanodot diameter. Snapshots corresponds to the diameter of 20 nm, 40 nm, and 60 nm with $p$ = 0.18. (c)(d) Snapshots of the magnetization dynamics with diameter of 60 nm. }	
\end{figure}

\begin{figure}[h]     
	\includegraphics[width=12cm]{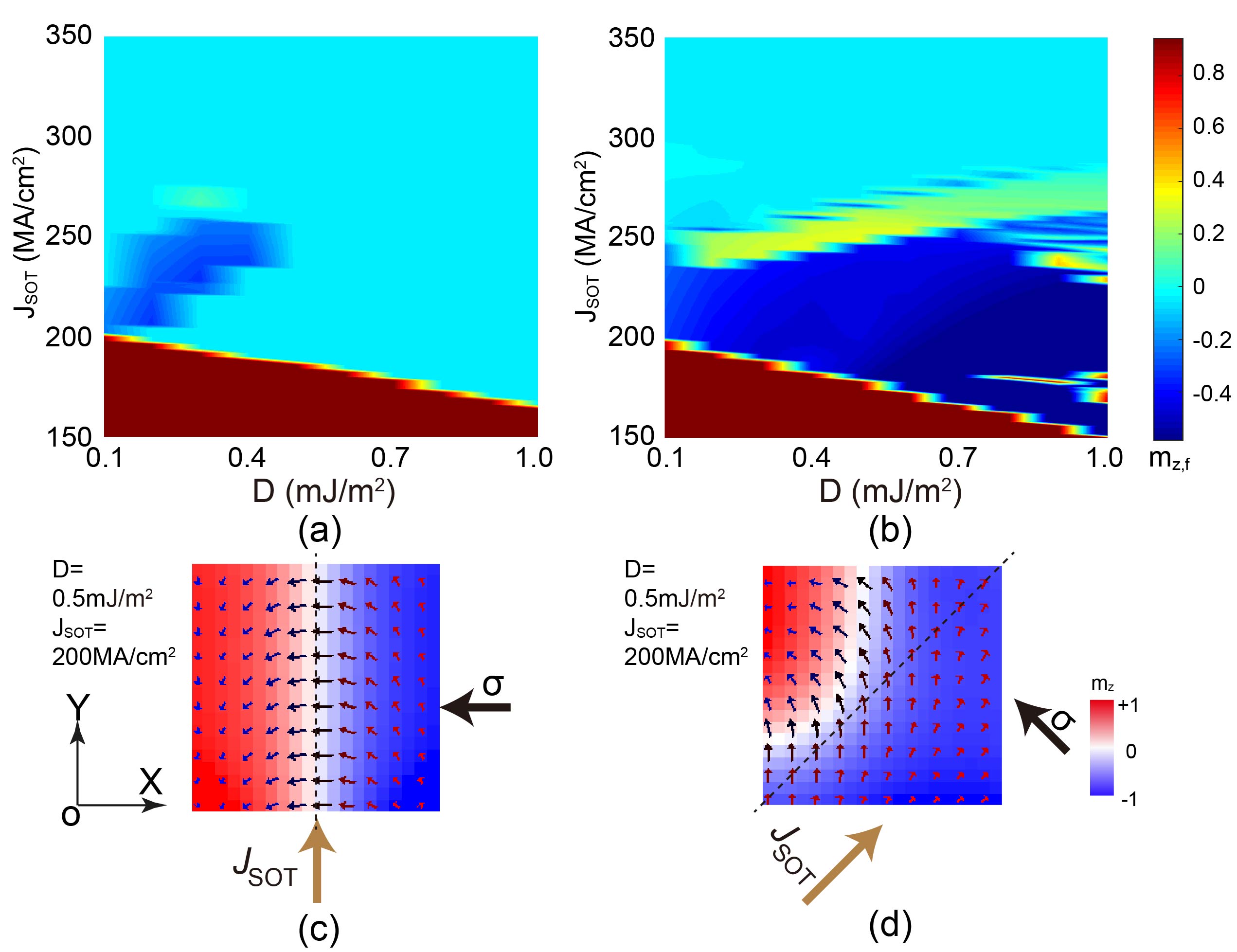}
	\centering
	\label{6}
	\caption{Phase diagrams of $m_{z,f}$ as a function of SOT current density and DMI magnitude (a) by applying +y current in a square film, (b) by applying titled current in a square film. (c) Typical metastable state corresponds to the light green region of (a). (d) Typical $bias$-$state$ corresponds to the dark blue region of (b).}	
\end{figure}

\newpage
\newpage

\end{document}